\documentclass[useAMS,usenatbib]{mn2e} 
\usepackage{graphicx,epsfig,psfig} 

\def\spose#1{\hbox to 0pt{#1\hss}} 
\def\simlt{\mathrel{\spose{\lower 3pt\hbox{$\mathchar"218$}} 
\raise 2.0pt\hbox{$\mathchar"13C$}}} 
\def\simgt{\mathrel{\spose{\lower 3pt\hbox{$\mathchar"218$}} 
\raise 2.0pt\hbox{$\mathchar"13E$}}}

\long\def\Ignore#1{\relax}

\newcommand{\degrees} {^\circ}

\title[Advanced Morphological Galaxy Classification]{Advanced Morphological Galaxy Classification:\break A Comparison of Observed and Simulated Galaxies}

\author[Hambleton et~al.]{K.M. Hambleton$^{1}$\thanks{Email: 
kmhambleton$@$uclan.ac.uk}, B.K. Gibson$^{1,2,3}$, C.B. Brook$^{1}$, G.S. 
Stinson$^{1,4}$, C.J. Conselice$^{5}$,\newauthor J. Bailin$^{4,6}$, H. Couchman$^{4}$ and J. Wadsley$^{4}$\\
$^{1}$Jeremiah Horrocks Institute, University of Central Lancashire, Preston, PR1~2HE, UK\\
$^{2}$Department of Astronomy \& Physics, Saint Mary's University, Halifax,
Nova Scotia, B3H~3C3, Canada\\
$^{3}$Monash Centre for Astrophysics, School of Mathematical Sciences, 
Monash University, Clayton, VIC, 3800, Australia\\
$^{4}$Department of Physics and Astronomy, McMaster University, 1280 Main Street West, Hamilton, Ontario, L8S 4M1, Canada\\
$^{5}$University of Nottingham, School of Physics \& Astronomy, Nottingham, NG7~2RD, UK\\
$^{6}$Astronomy Department, University of Michigan, 830 Dennison Bldg., 500 Church St., Ann Arbor, MI 48109-1042\\}

\begin{document}
\date{Accepted} 
\pagerange{\pageref{firstpage}--\pageref{lastpage}} \pubyear{2011} 

\maketitle 
\label{firstpage}

\begin{abstract}
Encoded within the morphological structure of galaxies are clues related 
to their formation and evolutionary history.  Recent advances pertaining 
to the statistics of galaxy morphology include sophisticated measures of 
concentration (C), asymmetry (A), and clumpiness (S). In this study, 
these three parameters (CAS) have been applied to a suite of simulated 
galaxies and compared with observational results inferred from a sample 
of nearby galaxies.  The simulations span a range of late-type systems, 
with masses between $\sim$10$^{10}$~M$_\odot$ and 
$\sim$10$^{12}$~M$_\odot$, and employ star formation density thresholds 
between 0.1~cm$^{-3}$ and 100~cm$^{-3}$. We have found that the 
simulated galaxies possess comparable concentrations to their observed 
counterparts. However, the results of the CAS analysis revealed that the 
simulated galaxies are generally more asymmetric, and that the range of 
clumpiness values extends beyond the range of those observed. Strong 
correlations were obtained between the three CAS parameters and colour 
(B-V), consistent with observed galaxies. Furthermore, the simulated 
galaxies possess strong links between their CAS parameters and Hubble 
type, mostly in-line with their observed counterparts.
\end{abstract}


\section{Introduction} 
\label{sec:intro}

Classification in astronomy has led to many advances which have revealed 
important information about the Universe in which we live. For example, 
the classification of stars by their colours and brightnesses in the 
Hertzsprung-Russel (HR) diagram led to an understanding of stellar 
structure and evolution.  The HR diagram continues to be employed as a 
means to to determine the ages and metallicites of stellar populations, 
whether they are simple stellar populations, as in globular clusters, or 
in composite populations, such as dwarf galaxies \citep{Dolphin2003, 
Monelli2010, Williams2010}.  Even when it is impossible to resolve 
stars, entire galaxies are classified by their colours and magnitudes. 
However, galaxies display resolved structure and morphology that stars 
do not, and using this morphology can reveal much about the evolution of 
galaxies.

Early methods used for the classification of galaxies were based on 
morphology.  The \citet{Hubble1926} Classification System is a method 
used to categorise galaxies by their morphology, and although it has 
been the dominant morphological tool since the mid-1920s, over time it 
has proven to be deficient in several areas. A prominent disadvantage to 
classifying galaxies based on their morphological features alone is that 
it is subjective with respect to distance (or resolution) and 
inclination. Furthermore it has led to the grouping of a wide range of 
asymmetric galaxies as simply ``irregular''.

As both ground- and space-based imaging has improved, and we probe 
higher redshifts, it has become apparent that classifying galaxies as 
``irregular'' means ignoring a vast amount of morphological information. 
\citet{deVaucouleurs1959} attempted to improve upon the 
\citet{Hubble1926} Classification System by introducing ``later types'' 
as a sub-class of galaxies; however, this system still refers to local, 
axisymmetric galaxies as reference points, and, again, was based solely 
on morphology.

\citet{Morgan1957} noted a correlation between the spectra of galaxies 
and the dominance of their central bulge component.  \citet{Morgan1958} felt that a shortcoming of the Hubble classification scheme was that it was not a strong indicator of 
spectral class. In order to rectify this,  \citet{Morgan1958} 
 devised a galaxy classification system based upon the central light 
concentration compared to its overall light distribution. This 
concentration classification scheme involved analysing the flux of 
galaxies to determine the degree of concentration of the central 
component.  The purpose of this scheme was to study the 
colour-concentration correlation in an attempt to identify galactic 
evolutionary states.  Due to the concentration parameter's direct 
relation with spectral class and type of stellar population, the 
concentration parameter was also found to be indicative of formation 
history and properties, such as velocity dispersion, galaxy size, 
luminosity, and (more recently) central black hole mass 
\citep{Graham2001}.

Extending Morgan's pioneering work, \citet{Conselice2000b} proposed a 
new statistical measure of asymmetry, in conjunction with a revised 
parametrisation of Morgan's concentration index.  This classification 
system was based on the desire to classify galaxies quantitatively at a 
range of redshifts.  Asymmetry was defined by rotating a galaxy 
180$\degrees$, about a central axis and finding the absolute sum of the 
normalised residuals.  This method measures the high frequency structure 
within each galaxy, whilst also considering the symmetry. The measure of 
asymmetry focuses primarily on morphological shape, and has a strong 
correlation with colour, congruent with Morgan's concentration 
parameter. As such, this property has a correlation with both star 
formation and merger history.

\citet{Conselice2003} introduced one additional morphological measure 
sensitive to high spatial frequency clumpiness. This parameter is 
defined by comparing a galaxy to a smoothed image of itself.  In doing 
so, the clumpiness parameter quantifies galaxy morphology in a manner 
which reflects star forming regions and evolutionary history.  This 
parameter distinguishes between varying types of galaxies and thus, 
partially, resembles Hubble's classification system.  However, the 
clumpiness parameter quantifies galaxy morphology with respect to 
intrinsic characteristics and not morphology alone.

\citet{Conselice2003} combined the high spatial frequency clumpiness 
parameter with both the asymmetry index and the concentration index to 
complete what is now called the CAS classification system. Underlying 
the CAS system's empirical nature lies the fundamental physics which 
governs its individual C, A, and S, components.  The CAS system 
highlights the intrinsic nature of galaxies by considering their light 
distributions and spectral class.

The observable properties of galaxies are a function of their merger histories,
masses and environments, each of which are reflected in the CAS system. It is fitting to apply this new classification method to the theoretically predicted properties of galaxies and use it as a gauge to determine where our current understanding of galaxy formation and evolution is correct and where it is deficient. Galaxy simulations are an integral tool that can help interpolate
between known stages of formation history, and thus further our understanding of
galaxy evolution.

Within the observational community, the CAS system has become one of the most
widely-used measures of galaxy morphology \citep{Hern2008, Bertone2009}. Since the
completion of the combined CAS system, it has been applied to numerous galaxies
including the 113 galaxies observed by \citet{Frei1996}. However, it has not yet
been applied systematically to high-resolution computational galaxy simulations as
a method of comparison between observations and simulations. In this work, we
develop a software package patterned on the \citet{Conselice2000b} and
\citet{Conselice2003} CAS system, calibrated on an optimised training set, and
applied to high resolution galaxy simulations. We show that the simulated galaxies exhibit both similarities and differences to real galaxies, and highlight the
intrinsic physical process that are responsible for the differences.

In \S\ref{SimSamp} we describe our simulated galaxy samples. In 
\S\ref{CAS} we discuss the three structural parameters of the CAS 
system, Concentration (C), Asymmetry (A)and Clumpiness (S). We then 
compare the simulations with the empirical \citet{Frei1996} sample of 
galaxies in \S\ref{Comparison}. The results are discussed in 
\S\ref{Discuss} and conclusions drawn in \S\ref{Concl}.

\section{Simulated Galaxy Sample}
\label{SimSamp}

\begin{figure}
\centering
\vspace{5.2cm}
\includegraphics{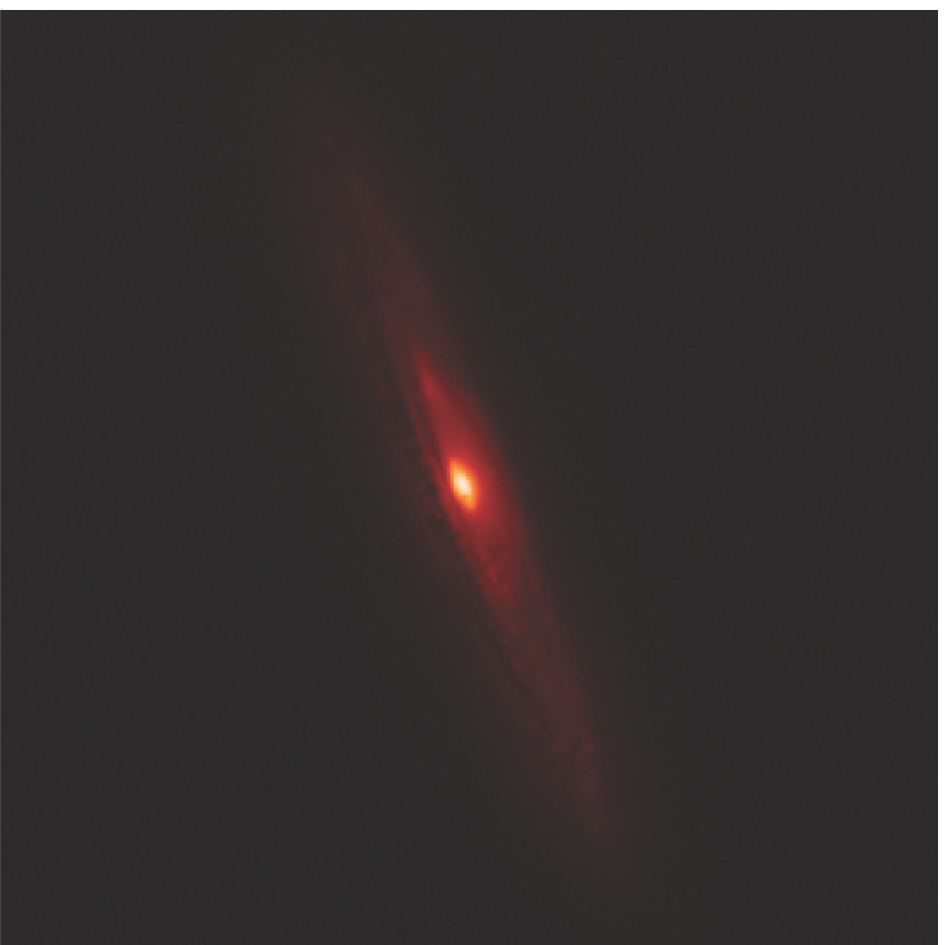}
\includegraphics{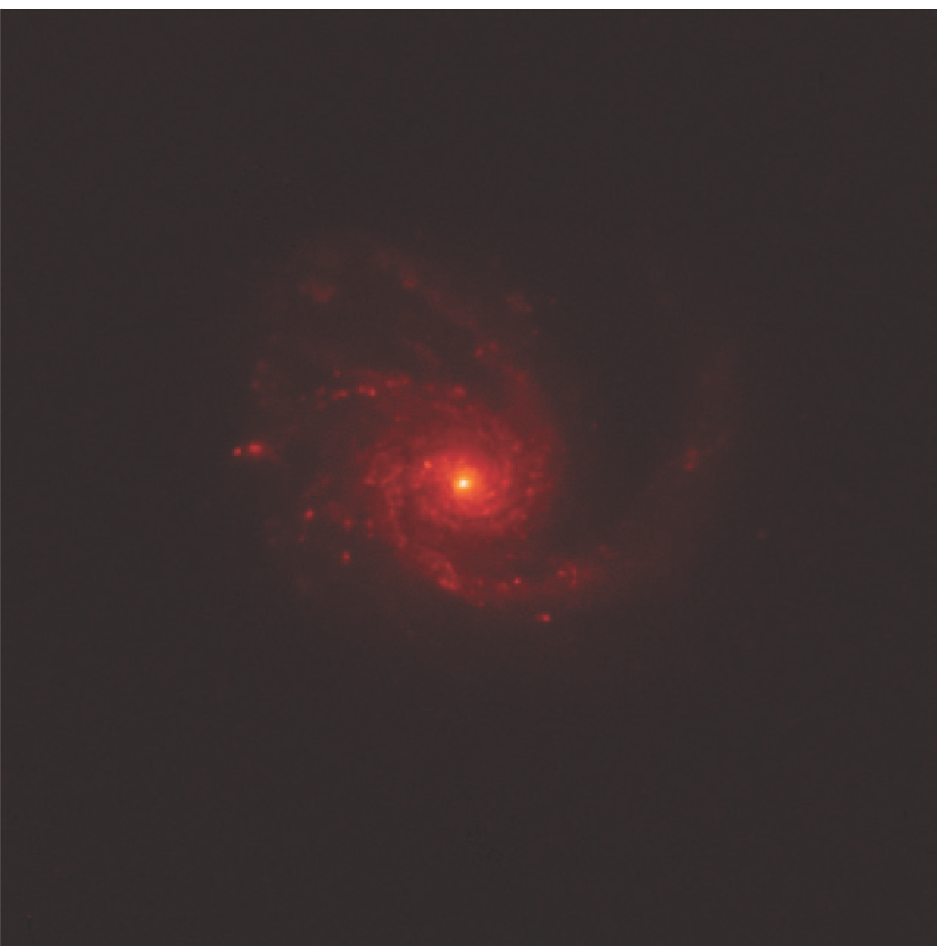}
\caption{\small {\it Left panel}: NGC~4216, an edge-on disk galaxy from the Frei et~al. (1996) sample in the SDSS $r$-band. {\it Right panel}: NGC~4254, a face-on disc from the Frei et~al. (1996) sample in the SDSS $r$-band.}
\label{Real} 
\end{figure}

\begin{figure}
\centering
\vspace{5.2cm}
\includegraphics{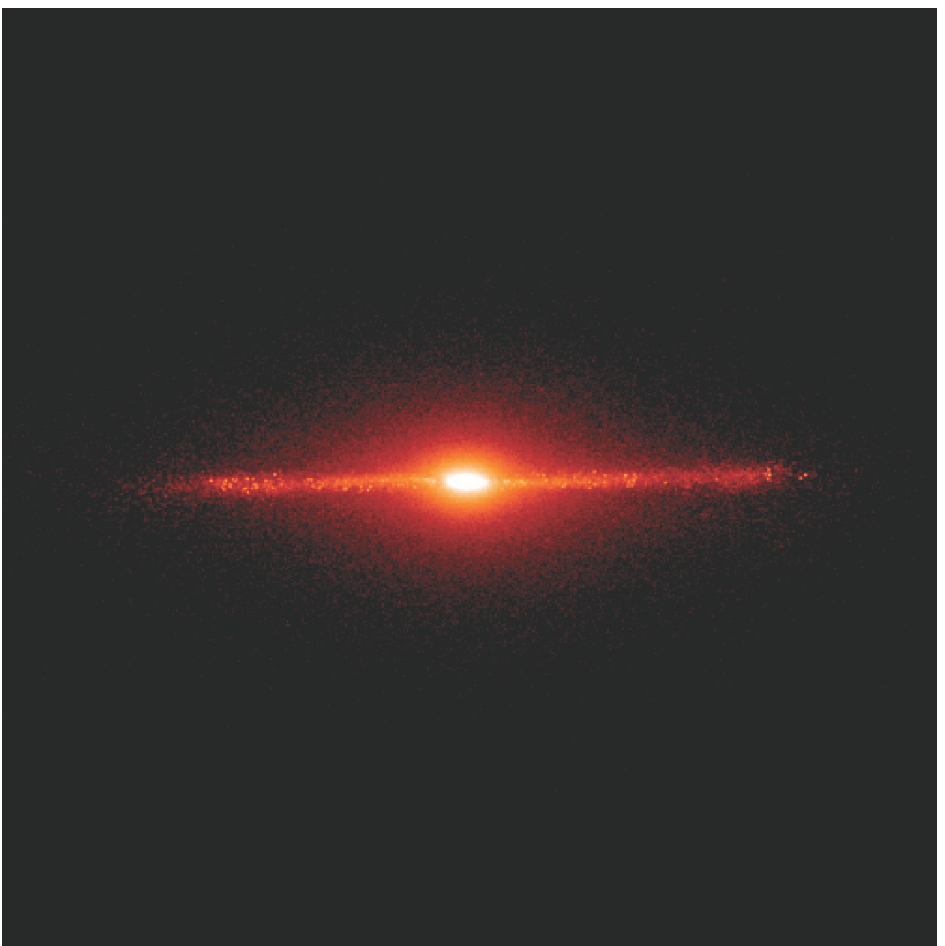}
\includegraphics{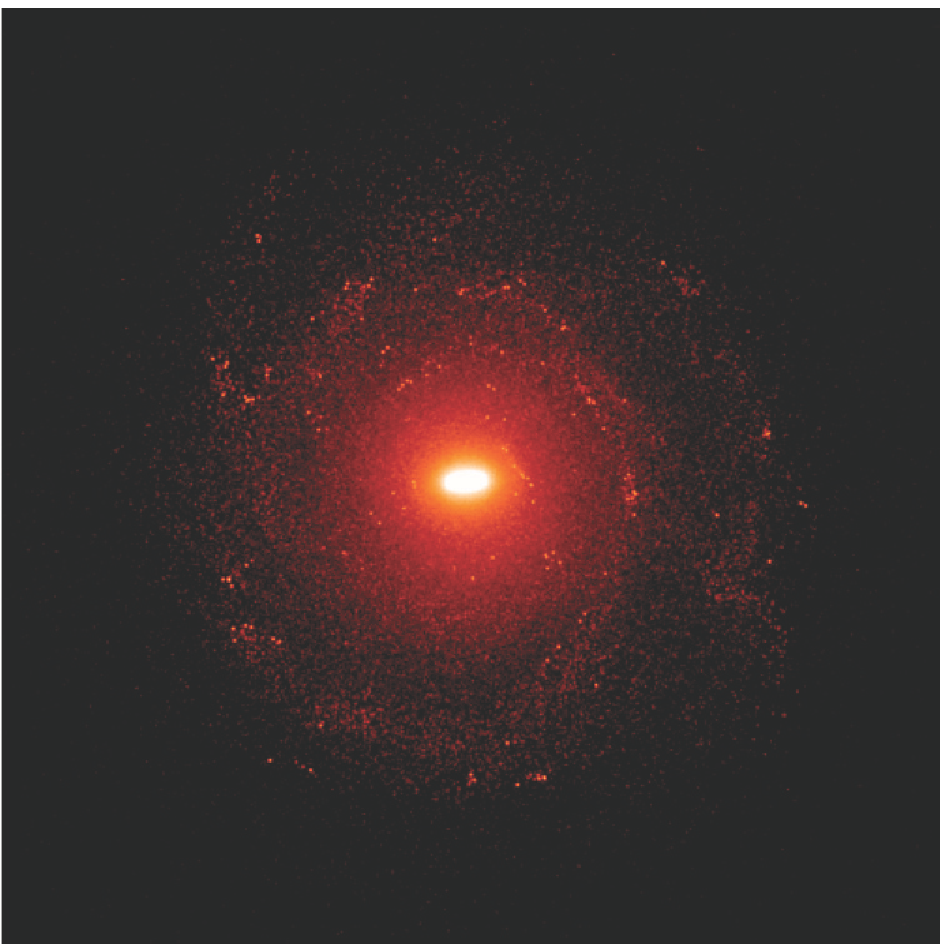}
\caption{\small {\it Left panel}: An image of g1536, a disc galaxy from the MUGS sample, as viewed edge-on in the SDSS $r$-band. {\it Right panel}: g1536, a disk galaxy from the MUGS sample, as viewed face-on in the SDSS $r$-band.}
\label{Stinson} 
\end{figure}

\begin{figure}
\centering
\vspace{5.2cm}
\includegraphics{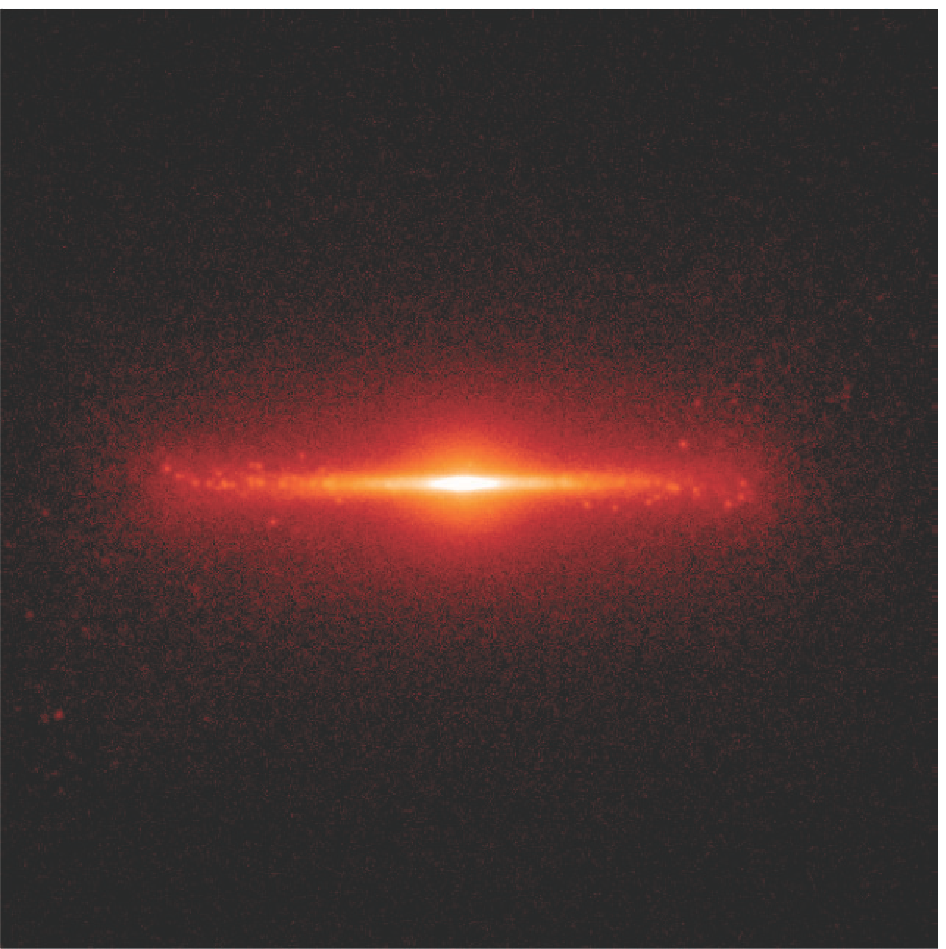}
\includegraphics{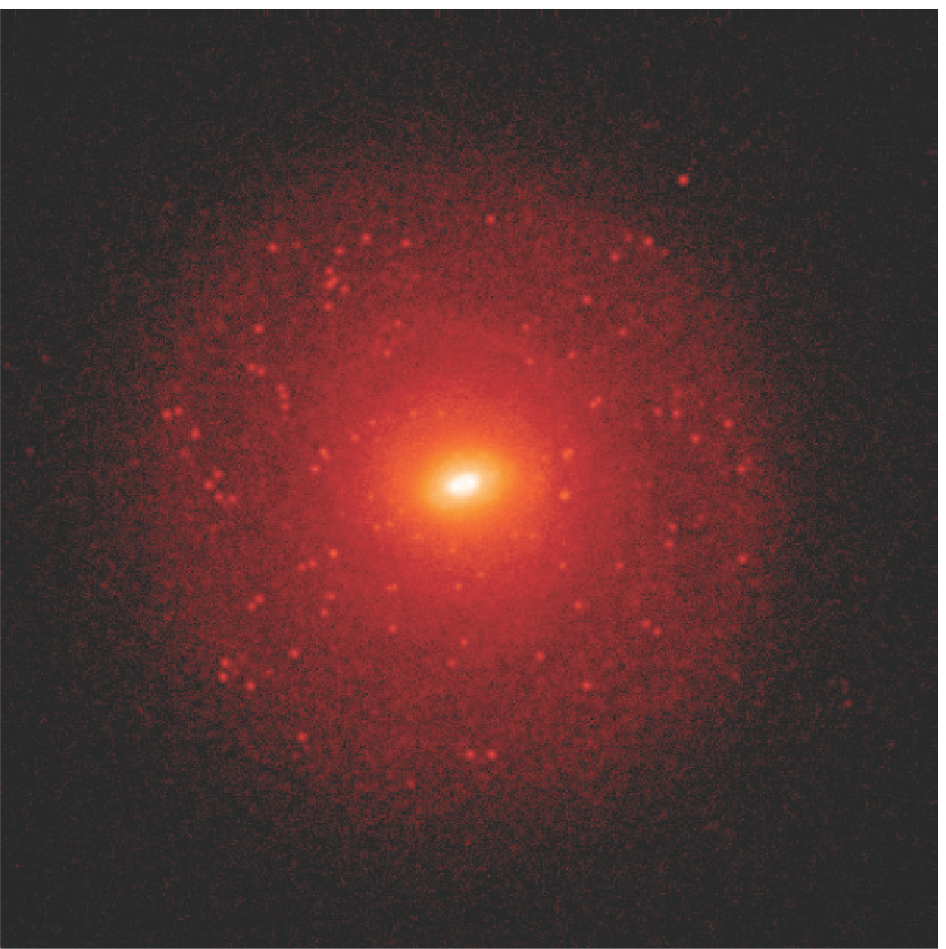}
\caption[MW]{\small {\it Left panel}: An image of MW, a disc galaxy from the UW sample, as viewed edge-on in the SDSS $r$-band. {\it Right panel}: An image of MW, a disc galaxy from the UW sample, as viewed face-on in the SDSS $r$-band.}
\label{UW} 
\end{figure}

\begin{figure}
\centering
\vspace{5.2cm}
\includegraphics{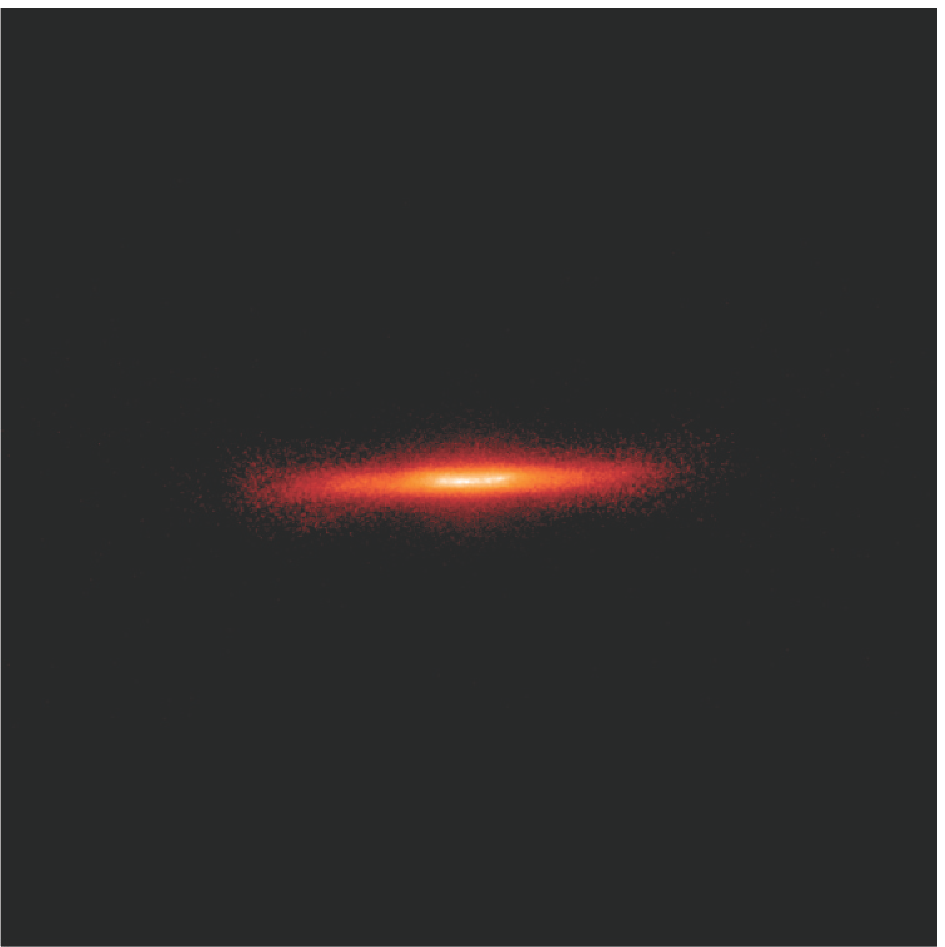}
\includegraphics{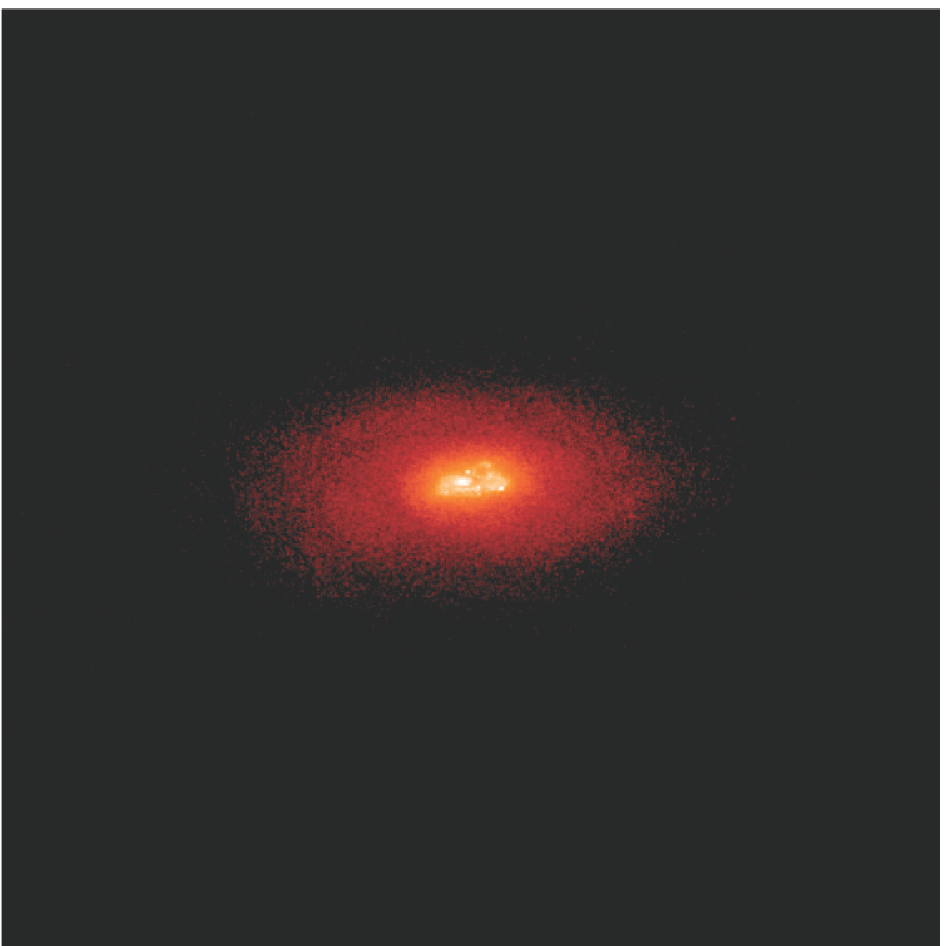}
\caption[DG3]{\small {\it Left panel}: An image of DG3, a disc galaxy from the Dwarf sample, as viewed edge-on in the SDSS $r$-band. {\it Right panel}: An image of DG3, a disk galaxy from the Dwarf sample, as viewed face-on in the SDSS $r$-band.}
\label{Dwarf} 
\end{figure}

To compare with the morphological parameters inferred from observed 
galaxies, we used three samples of simulated galaxies.  Each sample was 
generated using the gravitational N-Body + smoothed particle 
hydrodynamics (SPH) code \textsc{gasoline} \citep{Wadsley2004}. We 
briefly outline the simulations here, but refer the reader to the 
detailed descriptions found in \citet{Brooks2009}, for what will 
henceforth be called the ``UW sample'', \citet{Governato2010}, for the 
``Dwarf sample'', and \citet{Stinson2010} for the ``MUGS (McMaster Unbiased Galaxy Simulations) sample''.

Regardless of the aforementioned `sample' from which a specific 
simulation was drawn, each individual ``zoom-style'' simulation was run 
within a WMAP-3 $\Lambda$CDM cosmological framework \citep{Spergel2007} 
and consisted of a central high resolution region (centred on the target 
halo) embedded within a lower resolution cosmological volume.  Star 
formation and supernova feedback was computed using the recipes outlined 
by \citet{Stinson2006}, with the primary difference between the three 
samples being the adopted star formation density threshold ($n_{\rm 
th}$). For the UW sample, the canonical $n_{\rm th}$ of 0.1~cm$^{-3}$ 
was employed; for the MUGS galaxies, $n_{\rm th}$ was increased to 
1~cm$^{-3}$; for the Dwarf galaxy sample, $n_{\rm th}$=100~cm$^{-3}$ was 
adopted. The latter simulations represent the first successful 
realisations of essentially bulgeless disc galaxies (Governato et~al. 
2010). The masses of the MUGS and UW samples are comparable to that of 
the Milky Way ($\sim$10$^{12}$~M$_\odot$) and, therefore, also the 
representative massive galaxies in the \citep{Frei1996} sample, while 
those of the Dwarf sample are of the order $\sim$10$^{10}$~M$_\odot$ 
(more akin to the Small Magellanic Cloud). We include one 
additional dwarf (labelled DG1LT) which was simulated using the same 
parent dark matter halo as one of the canonical dwarf simulations (DG1), 
but employed a lower star formation threshold of 0.1~cm$^{-3}$ (similar 
to the UW sample).

We used \textsc{SUNRISE} \citep{Jonsson2006}, a Monte Carlo radiative transfer
code, to generate \it a posteriori\rm, mock images of the simulations,
incorporating the effects of dust attenuation and spanning a range of viewing
angles from edge- to face-on perspectives, as listed in Table~\ref{tab:CAS}.

All galaxies are presented at inclinations of both 0$\degrees$ and 90$\degrees$,
and in most cases also at a uniform distribution of intermediate inclinations.
Within this distribution there are some correlations due to the intrinsic features
of each galaxy that are maintained at all inclinations; e.g. in
Figure~\ref{ClumpAsym}, DG2 is clumpy at all inclinations. We view each galaxy
from a range of inclinations, which allows us to compare them to observed
galaxies, which are viewed at a range of inclinations. Furthermore, it increases
the size of the sample given the limited number of galaxy simulations available at
sufficient resolution. Subsequently we have multiple copies of the same physical
system, thus the images are not truly independent and care should be taken not to
over-interpret any statistical conclusions. In Figures 1 to 4, SDSS $r$-band
images of representative edge-on and face-on galaxies drawn from the observational
(Figure~1) and simulated (Figures 2-4) samples are shown.

\section{The CAS Structural Parameters}
\label{CAS}

The three CAS parameters were recreated after \citet{Conselice2000b} and 
\citet{Conselice2003}, and calibrated on the \citet{Frei1996} sample of 
observed nearby galaxies in the 'r' and 'R' bands. Prior to the computation of 
each parameter, each galaxy was background subtracted by initially 
taking the median count value of a section at the edge of each galaxy's 
CCD frame, and subtracting this from each pixel within the array. The 
surface brightness profile was then inspected to ensure that the correct 
level of background was removed.

Prior to computing the CAS indices, a characteristic radius for each 
galaxy was determined. \citet{Conselice2000b} recommends the use of the 
characteristic $\eta$ radius which is 1.5~r$_{\rm Petrosian}$. The Petrosian radius is defined as the radius $r$ at which the ratio of the local surface brightness (in an annulus from $0.8r$ to $1.25r$) to the mean surface brightness within $r$ is 0.2. This is found by calculating the aforementioned ratio for radii within an annulus 0.8\,r to 1.25\,r.

\subsection{Concentration (C):}

The concentration parameter is a measure of how concentrated (or 
diffuse) the central bulge component is with respect to the total flux 
of the galaxy. Concentration has a strong correlation with colour, and 
is related to other intrinsic features such as velocity dispersion and 
mass \citep{Graham2001}. The concentration of a galaxy is determined by 
locating the radii of the circular apertures that contain 20\% 
($r_{0.2}$) and 80\% ($r_{0.8}$) of the total flux of the galaxy, and 
then taking their ratio:

\begin{equation}
C = 5\log \frac{r_{0.8}} {r_{0.2}} 
\end{equation}

\subsection{Asymmetry (A):}

The asymmetry parameter (A) is a measure of how symmetric (or 
asymmetric) a galaxy is with respect to the total flux of the galaxy. 
Galactic asymmetries are predominantly caused by star formation, mainly 
in underlying structure such as spiral arms. Asymmetry is also a strong 
indicator of merger history, as demonstrated by \citet{Conselice2003}. 
The asymmetry parameter initially involves locating the centre of a 
galaxy image, which is defined as the global minimum of asymmetry. The 
centre was obtained using the centering algorithm specified by 
\citet{Conselice2000b}, which involves an iterative trial-and-error 
process. An initial guess for the centre of the galaxy is made and the 
asymmetry for the specified pixel, and the surrounding eight pixels is 
determined. The pixel corresponding to the lowest asymmetry is then 
redefined to be the centre. In order to locate the global asymmetry 
minimum, which corresponds to the centre, it is required that the 
process be repeated until the central pixel is also the pixel with the 
lowest value of asymmetry.

The asymmetry itself is calculated by rotating the image 180$\degrees$ 
about the defined central point and subtracting the rotated image from 
the initial image, in order to determine the absolute value of the 
residuals. Following this, the residuals are divided by the quantity 
corresponding to the summation of all the pixels in the galaxy.  
Subsequently, a similar method is then applied to a small section, taken 
from a corner of the image which contains background only. This quantity 
is then multiplied by a scale factor so that it is comparable to the 
size of the galaxy image. This quantity is divided by the summation of 
the total flux of the original galaxy image. This is done in order to 
determine the asymmetry of the background so that it may be discounted 
from the galactic asymmetry. The calculation for the asymmetry is as 
follows:

\begin{equation} A = \frac{\sum 
I-I_{180}}{\sum I} - {\frac{\sum B-B_{180}}{\sum I}} 
\end{equation} 

\noindent
where {\it I} is the image flux, {\it I$_{180}$} is the rotated image 
flux, {\it B} is the background, and {\it B$_{180}$} is the rotated 
background.

\subsection{Clumpiness (S):}
\label{Clump}

The measure of high-spatial frequency clumpiness (S) is directly related 
to a galaxy's star formation and high frequency spatial power.  The 
clumpiness is defined as the patchiness of the high frequency light 
distribution within a galaxy \citep{Conselice2003}. It is computed by 
subtracting a degraded-resolution version of the galaxy image from the 
original galaxy image.  As with the asymmetry parameter the method is repeated on a section of sky background 
in order to isolate the fraction of the signal that is due to the galaxy.  Again, a corner of the image 
containing background only is selected, replicated, and smoothed, so 
that an identical procedure may be carried out on both the background 
and the galaxy image.

As the central bulge component is a prominent feature in the simulated 
galaxies, it contributes to a large portion of the high frequency 
structure and hence clumpiness of the simulated galaxies. As it is 
already a known issue that simulated galaxies tend to have large central 
bulge components, with respect to their observed counterparts, it was 
decided that the standard central region specified by Conselice (2003) 
would be extended to include the total central bulge component of each 
galaxy. Furthermore, the inferred total flux of the galaxies would also 
exclude the central region, in order to avoid any bias due to the 
varying sizes of the bulge components of the galaxies.  The size of the 
central bulge component was determined from the light profile of the 
galaxy by considering the radius at which the profile changed from 
ln($I$) ${=}$ r$^{1/4}$, to exponential decay. The image was then 
examined to ensure that the correct area had been removed, prior to the 
computation of the clumpiness parameter.

The calculation  for the clumpiness is as follows:

\begin{equation}
S = 10 \times \left(\frac{\sum I-I_{S}}{\sum I} - {\frac{\sum B-B_{S}}{\sum I}}\right)
\label{eqn:clump}
\end{equation}

\noindent
where {\it I} is the image flux, {\it I$_{s}$} is the smoothed image 
flux, {\it B} is the background, and {\it B$_s$} is the smoothed 
background. Further details on the derivation of the clumpiness 
parameter are provided in Appendix~A.

\begin{table}
\caption{CAS parameters for simulated galaxies, including values 
obtained at different inclinations. The names of the galaxies were 
defined by the teams that simulated them (see text for details); the
specific names are arbitrary for the purposes of this study.}
\begin{tabular}{c|c|c|c|c}
\hline
\multicolumn {5}{l}{UW Sample}\\\hline
Galaxy & Inclination ($\degrees$) & C(r) & A(r) & S(r)\\\hline
Gal1   & 0      & 3.49    & 0.351  & 0.485\\   
       & 90     & 3.26    & 0.312  & 0.680\\   
h277   & 0      & 4.96    & 0.215  & 0.126\\   
       & 90     & 3.82    & 0.161  & 0.508\\   
MW1024 & 0      & 3.97    & 0.180  & 0.181\\   
       & 90     & 3.65    & 0.180  & 0.357\\ 
\hline
\multicolumn {5}{l}{Dwarf Sample}\\
\hline
Galaxy & Inclination ($\degrees$)& C(r) & A(r) & S(r)\\\hline  
DG1    & 0      & 3.38    & 0.385  & 0.378\\   
       & 45     & 3.38    & 0.232  & 0.453\\    
       & 90     & 3.47    & 0.369  & 0.667\\
DG1LT  & 0      & 2.80    & 0.267  & 0.259\\
       & 30     & 3.01    & 0.281  & 0.253\\
       & 45     & 2.87    & 0.275  & 0.303\\
       & 60     & 3.18    & 0.260  & 0.380\\
       & 90     & 4.06    & 0.231  & 0.761\\
DG2    & 0      & 3.66    & 0.574  & 0.217\\
       & 30     & 3.17    & 0.570  & 0.193\\
       & 45     & 3.19    & 0.568  & 0.228\\
       & 60     & 3.18    & 0.552  & 0.197\\
       & 90     & 3.17    & 0.376  & 0.699\\
DG3    & 0      & 3.28    & 0.315  & 0.139\\
       & 30     & 3.31    & 0.317  & 0.126\\
       & 45     & 3.29    & 0.314  & 0.134\\
       & 60     & 3.22    & 0.311  & 0.177\\
       & 90     & 3.48    & 0.200  & 0.664\\

\hline
 \multicolumn {5}{l}{MUGS Sample}\\
\hline
Galaxy & Inclination ($\degrees$) & C(r) & A(r) & S(r)\\\hline
g1536   & 0     &  4.51    & 0.158 & 0.078\\
       & 45    &  4.46    & 0.174 & 0.084\\
       & 90    &  4.11    & 0.123 & 0.201\\
g21647  & 0     &  3.49    & 0.162 & 0.104\\
       & 45    &  3.31    & 0.150 & 0.061\\
       & 90    &  3.49    & 0.173 & 0.314\\   
g22437  & 0     &  3.71    & 0.240 & 0.258\\  
       & 30    &  3.49    & 0.193 & 0.186\\
       & 45    &  3.34    & 0.209 & 0.159\\
       & 60    &  3.31    & 0.222 & 0.146\\
       & 90    &  3.45    & 0.209 & 0.192\\  
g24334  & 0     &  3.57    & 0.235 & 0.362\\ 
       & 30    &  3.44    & 0.235 & 0.424\\
       & 45    &  3.44    & 0.230 & 0.406\\
       & 60    &  3.36    & 0.215 & 0.393\\
       & 90    &  3.54    & 0.246 & 0.457\\   
g25271  & 0     &  3.63    & 0.106 & 0.007\\ 
       & 45    &  3.63    & 0.086 & 0.007\\
       & 90    &  4.06    & 0.148 & 0.100\\   
g3021   & 0     &  3.71    & 0.206 & 0.277\\ 
       & 30    &  3.76    & 0.216 & 0.324\\
       & 45    &  3.84    & 0.235 & 0.393\\
       & 60    &  3.91    & 0.234 & 0.407\\
       & 90    &  4.03    & 0.209 & 0.419\\   
g5664   & 0     &  3.16    & 0.146 & 0.359\\ 
       & 45    &  2.93    & 0.170 & 0.330\\
       & 90    &  3.39    & 0.142 & 0.726\\   
g7124   & 0     &  4.51    & 0.077 & 0.000\\ 
       & 45    &  4.00    & 0.099 & 0.029\\
       & 90    &  4.22    & 0.142 & 0.001\\\hline
\end{tabular}
\label{tab:CAS}
\end{table}

\section{Comparison of observations with simulations}
\label{Comparison}

The CAS parameters form a well-established system that has been used on 
occasion to analyse galaxy simulations, although not in any systematic 
manner.  For example, \citep{Conselice2006} applied the system to dark 
matter merger-remnant simulations, and \citep{Lotz2008} extended 
this to the analysis of equal-mass gas-rich disc merger remnants.  Our 
unique contribution to this area is, for the first time, an application 
to a suite of simulations realised within a 
fully cosmological framework. Such a comparison should enable underlying 
physical differences to be revealed and allow for scaling relations to 
be tested using simulations in a new and unique manner. The main purpose 
of this paper is to consider the morphological details of the galaxies 
produced in simulations, and uncover new information regarding the 
differences between observed and simulated galaxies. To avoid simply 
re-identifying known issues and biases, we have modified slightly the 
underlying CAS parameters, as explained now.

\subsection{Concentration (C):}

The overproduction of the centrally-concentrated bulge component of most 
simulated disc galaxies (e.g., Stinson et~al. 2010, and references 
therein; c.f., Governato et~al. 2010), is a long-standing problem 
associated with cosmological simulations of galaxy formation. By 
definition, the concentration parameter is a proxy for the dominance of 
the central region, simulation or not, although consideration of the 
surface brightness profiles suggested that the basic definition not be 
altered here.  Instead, we concentrate here on the relationship between 
the size of the central component and disk of each galaxy, and compare 
this with observations.

Another discrepancy between observed and simulated galaxies is their noise content as noise does not overlay the simulated galaxies. Unlike the other CAS parameters, the 
measure of concentration does not include a background subtraction 
stage. This is due to the negligible effects of noise on the measure of 
concentration. For this reason, no adjustments were made to the 
concentration parameter regarding background noise.

\subsection{Asymmetry (A):}

The measure of asymmetry considers the residuals produced when a galaxy 
image is subtracted from the same image that has been rotated by 
180$^\circ$. The central bulge component of a galaxy is generally 
symmetric, and for this reason no adaptation was required to account for 
the excessive central bulge component of the simulated galaxies. However, it was discovered that the effects of noise were critical, as the background component was comparable in magnitude to the component due to the galaxy itself. It was imperative to ensure that this aspect of the algorithm was 
correct, as simulated galaxies contain essentially zero background noise 
compared to their observed counterparts, the latter of which contain varying 
amounts of noise depending upon exposure time, sky background, and 
distance. For this reason, any discrepancies concerning the computation 
of the background would lead to an inaccurate comparison.

In order to test this hypothesis, an artificial array of noise was 
created, and the galaxies were analysed both before and after the application 
of this noise array. We replicate the signal-to-noise values of the Frei et al. (1996) sample by first pixelating the simulated images to the same resolution, then scaling the flux to be comparable to that in the images of the real galaxies, and finally adding a normally-distributed noise array of sufficient amplitude to reproduce the signal-to-noise of the observed galaxies in Frei at al. (1996). Furthermore, \citet{Conselice2000b} demonstrated that accurate asymmetry calculations can only be performed, without corrections, on galaxies with signal-to-noise greater than 100. For this reason, each galaxy was tested individually to ensure that the application of these characteristic noise values led to signal-to-noise ratios in excess of 100.

\begin{figure}

\includegraphics[width=\hsize]{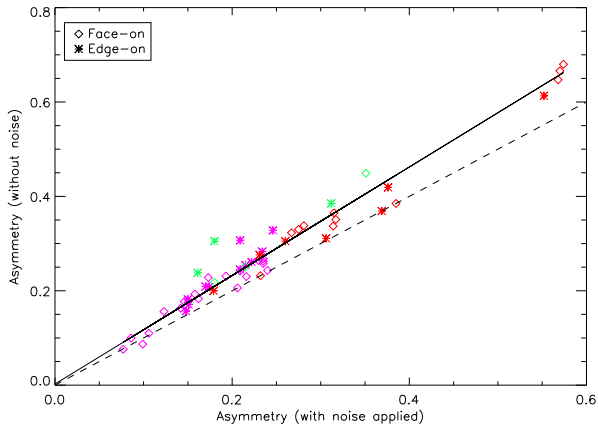}
\small\caption{Relationship between the asymmetry of simulated galaxies 
both before and after the application of noise. The solid line 
represents the least squares regression line for which the equation is 
$y = 1.15x - 0.002$, and the dashed line corresponds to the one-to-one 
relation. Here, the points are coloured to represent the different 
galaxy samples: MUGS = purple, UW = green and 
Dwarf = red. The choice of symbol corresponds to 
galaxy inclinations of 0 to 45 degrees (`face-on': diamonds) and 45 to 
90 degrees (`edge-on': asterisks).}
\label{AsymN} 
\end{figure}

A comparison of the asymmetry results derived from the simulated 
galaxies, both before and after the application of noise, can be seen in 
Figure~\ref{AsymN}. It was found that the inferred asymmetry values 
differed only marginally after the application of noise (in large part 
because of the stringent signal-to-noise limit imposed), although the 
impact varied for each galaxy and appeared to be more substantial for 
more asymmetric galaxies. Whilst there is some variation visible in 
Figure~\ref{AsymN} due to the application of noise, it is not 
significant. However, to ensure that a fair comparison was made between 
observed and simulated galaxies, it was decided that noise would be applied 
to the simulated galaxies prior to the computation of the asymmetry 
parameter.

\subsection{Clumpiness (S):}

When initially computed, the central bulge component dominated the 
clumpiness calculation for the simulated galaxies, suggesting that the 
simulations were excessively clumpy with respect to their observed 
counterparts. Prior to calculating the clumpiness, \citep{Conselice2003} 
forces the central region, which corresponds to 0.05 times the Petrosian 
radius, to zero, to ensure that any unresolved region of the central 
bulge component is excluded. For the purpose of this study this region 
has been extended to encompass the total central bulge component of each 
galaxy, thus allowing for the comparison of observed and simulated galaxies 
outside the bulge region, which we know is overproduced in these 
simulations. By making this alteration, the clumpiness of the disk 
region only is calculated, excluding any high frequency structure 
associated with the central bulge component.

As the area of the central bulge differs for each galaxy, in order to 
avoid any bias due to bulge size, it was then decided that the central 
bulge component would also be excluded from the summation of the total 
flux which is present in the denominator of the equation for the 
computation of the clumpiness (recall, \S\ref{Clump}).

The application of noise had a negligible effect upon the values 
generated by the clumpiness algorithm. This suggests that the 
subtraction algorithm for the clumpiness parameter is more than adequate 
for minimising the impact of noise on its calculation. For this reason 
it was deemed unnecessary to apply noise to the simulated galaxies in 
the computation of the clumpiness.

\section{Discussion}
\label{Discuss}

For each simulation, surface brightness profiles, for a range of 
inclinations, were generated.  The simulations were `pixelated' to a 
scale corresponding to 100~pc/pixel, which, as shown by 
\citet{Conselice2000b}, appears adequate for CAS analysis. The 
\citet{Frei1996} sample of observed, nearby galaxies, used in this study, 
all possess resolutions comparable to (or better) than the same 
100~pc/pixel, and hence are again ideal for use with the CAS system. As 
all the galaxies used in this study are above the threshold value of 
$\sim$1~kpc/pixel, where the accuracy of the CAS parameters begins to 
decline \citep{Conselice2000b}, a comparison between simulated and observed 
galaxies can be made with a high degree of confidence.

The \citet{Frei1996} sample of galaxies contains only nearby high 
surface brightness galaxies. It is not meant to be a complete and/or 
unbiased compilation, as it mostly excludes irregular and low surface 
brightness dwarf galaxies, which contribute significantly to the 
population of local galaxies. Similarly, the sample of simulated 
galaxies is incomplete in the sense that it does not contain any 
irregular or elliptical galaxies, resulting in a restricted range of 
colours; this impacts particularly on the red end of the distribution, 
where we lack counterparts to the \citet{Frei1996} sample. This does 
place constraints on the ability to compare observed and simulated galaxies, 
and is considered during this analysis by incorporating Hubble type in many of the figures.

\subsection{Concentration}

\begin{figure}
\includegraphics[width=\hsize]{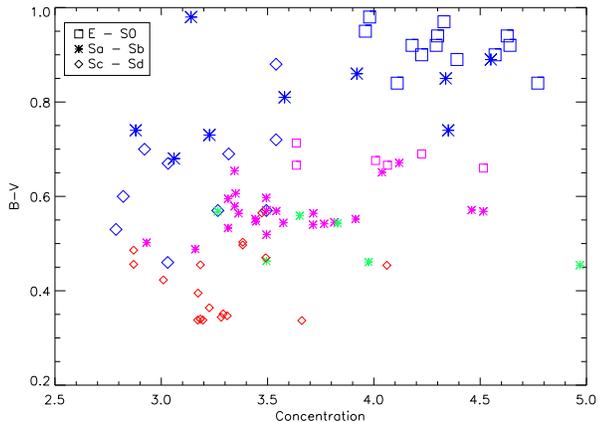}
\small\caption{Relationship between concentration and colour (B-V) for 
both observed and simulated galaxies. The points are coloured to represent 
the different galaxy samples: observed = blue (and slightly larger in size 
relative to the other symbols), MUGS = purple, UW = green, and Dwarf = 
red. The symbols correspond to the classical Hubble types, as noted in 
the inset.}
\label{ColourConc}
\end{figure}

In Figure~\ref{ColourConc}, both observed and simulated galaxies demonstrate 
a strong correlation between colour and concentration. This trend 
naturally arises for observed galaxies where early-type galaxies are 
redder in colour and also tend to be more centrally-concentrated, 
whereas later-types are bluer due to ongoing star formation and tend to 
be less centrally-concentrated. These values are well quantified by 
Hubble type and hence, for the observed galaxies, it can be seen that the 
different symbols that represent the various Hubble types are contained 
within well-defined regions. The visual appearance of a trend among the simulated galaxies comes more from the systematic offset between the Dwarf sample in red and the MUGS and UW samples in purple and green. While there is a trend within the MUGS sample of galaxies, it is not nearly as visually striking.

It is evident from Figure~\ref{ColourConc} that the simulated galaxies 
tend to be bluer than the observed galaxies, both in general and with 
respect to their Hubble type. The main cause of this blueward offset is 
the presence of excessive late-time star formation in the simulated 
discs. Furthermore, there are a lack of elliptical galaxies in the 
simulated sample, with the earliest galaxies being lenticular. However, 
in a relative sense, there is a definite relationship between colour and 
Hubble type, as expected.

The simulated galaxies in Figure~\ref{ColourConc} show a similar range 
of concentrations to those observed for real galaxies, and also appear 
to be divided into groups by Hubble type. Furthermore, the simulated 
galaxies demonstrate a positive linear trend that, although offset due 
to colour differences, is comparable to the trend demonstrated by the 
observed galaxies. Concentration is also related to surface brightness; it 
is expected that the low concentration values generated by the majority 
of the simulated dwarf galaxies is primarily due to their low surface 
brightness. Furthermore, the dwarf galaxies in this sample are all 
younger disc-like galaxies which are observed to be generally more 
diffuse. Other factors which influence the concentration parameter 
include the mass and velocity dispersion, both of which are low for the 
dwarf galaxy sample.

The MUGS sample of simulated galaxies in Figure~\ref{ColourConc} appears 
to extend to higher values of concentration than both the UW sample of 
Milky Way-like galaxies and the Dwarf sample. This is most likely due to 
the presence of earlier-type discs that have greater masses within the unbiased MUGS sample 
(c.f., with the UW sample, which were chosen a priori to more 
closely resemble later-type discs). Moreover, the MUGS sample of 
galaxies have a higher density threshold for star formation (1 
cm$^{-3}$), which could lead to preferential star formation in the 
denser (central) regions of the galaxy (Agertz, Teysier \& Moore 2010), 
although this behaviour is not seen in dwarf galaxy simulations 
(Governato et~al. 2010). The MUGS sample also has a greater average 
circular velocity (150 - 250 km/s), compared to the other simulated 
samples of galaxies, which is a parameter known to correlate with 
concentration \citep{Graham2001}.

\subsection{Asymmetry}

\begin{figure}
\includegraphics[width=\hsize]{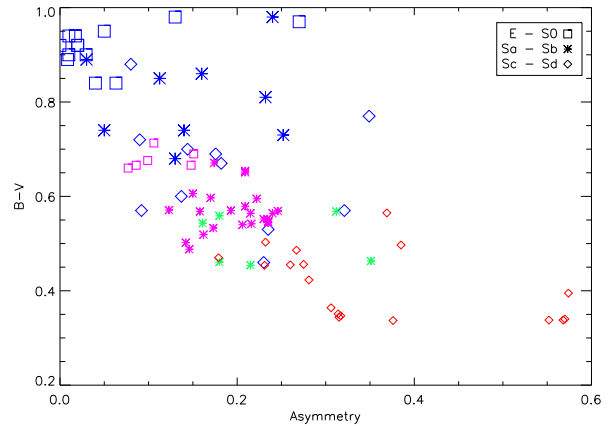}
\small\caption{Relationship between asymmetry and colour (B-V) for both 
the observed and simulated galaxies. Symbols and colours are the same as 
in Figure~\ref{ColourConc}.} 
\label{ColourAsym}
\end{figure}

In Figure~\ref{ColourAsym}, the observed galaxy sample demonstrates a strong 
anti-correlation between asymmetry and colour. This correlation stems 
from the bluer, younger galaxies generally containing spiral arms with 
dust lanes and star forming regions, and the older, elliptical galaxies 
generally being devoid of structure. The simulated galaxies follow an 
approximately identical trend to that of the observed galaxies, however, due 
to the bluer nature of the selected simulations, they form a blue 
extension of the observed galaxy trend. \citet{Conselice2003} showed 
that there was a distinct region in the bottom left corner of the 
colour-asymmetry plot where no galaxies were bluer and more symmetric. 
This region is well-defined by the simulations which demonstrate a very 
strong correlation between colour and asymmetry.

The earlier-types within the MUGS sample are redder in colour with 
respect to the other simulations, and tend to be more symmetric. Again 
there is a strong distinction between the various Hubble types, which is 
similar to the correlation demonstrated by the observed galaxy sample. 
However, due to the lack of redder simulated galaxies with low, 
approximately zero, asymmetry values, this trend does not extend to the 
upper left corner of Figure~\ref{ColourAsym}, where the redder, 
symmetric, galaxies reside in nature. Again, this can be traced to the 
ongoing late-time star formation in the earlier-type simulations which 
is not seen in nature. In Figure~\ref{ColourAsym} some of the dwarf 
simulated galaxies appear to be excessively asymmetric relative to the 
other simulations. The large variation of apparent flux levels within 
these systems is most likely the cause of this increased in asymmetry.

\begin{figure}
\includegraphics[width=\hsize]{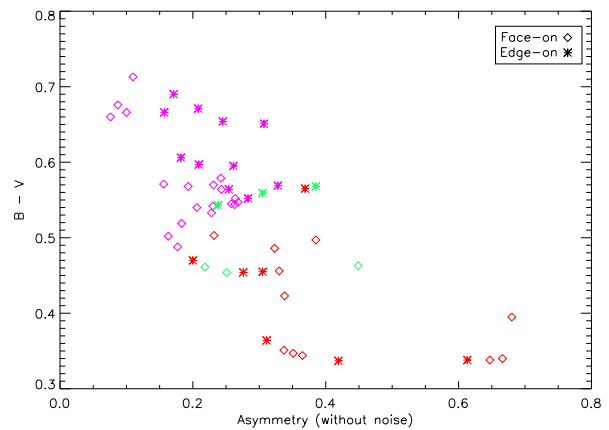}
\small\caption{Relationship between the asymmetry (A) of simulated 
galaxies, without the application of noise, and colour (B-V). Symbols 
and colours as in Figure~\ref{AsymN}.}
\label{ColourAsymN}
\end{figure}

\citet{Conselice2000b} showed that it is almost always the inclined 
galaxies that contribute to the scatter in the colour-asymmetry diagram and that the less inclined galaxies are systematically higher.
In Figure~\ref{ColourAsymN}, the edge-on and face-on galaxies have 
different symbols, and from this it can be concluded that while the 
inclined, observed galaxies do tend to be somewhat more significant 
outliers, in detail, the scatter of the simulated galaxies is not a 
function of inclination angle. Also, the MUGS and UW galaxies that are less inclined do appear to be systematically higher in the plot, as found for the observed galaxies.

Noise was applied to the simulated galaxies to ensure that a fair 
comparison was made when considering the asymmetry. The addition of a 
noise array to the simulated galaxy images had the benefit of allowing 
for a more accurate comparison with the observed galaxies. In 
Figure~\ref{ColourAsymN} the range of asymmetry values, without the 
addition of noise, can be seen. The absence of noise only appears to 
affect the outcome of the asymmetry parameter at greater asymmetry 
values by restricting the range of asymmetries.

\subsection{Clumpiness}

\begin{figure}
\includegraphics[width=\hsize]{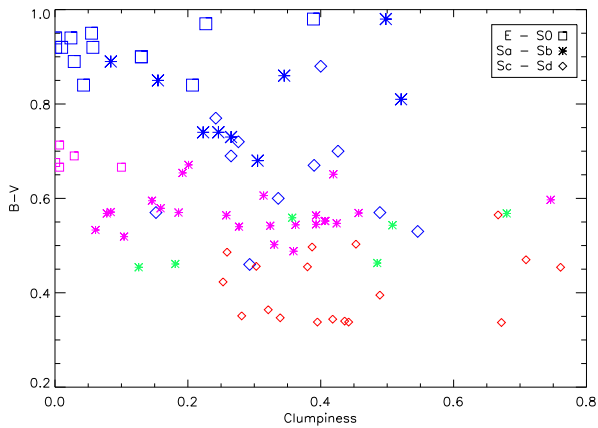}  
\small\caption{Relationship between clumpiness (S) and colour (B-V) for 
both the observed and simulated galaxies. Symbols and colours as 
in Figure~\ref{ColourConc}.}
\label{ColourClump}
\end{figure}

In Figure~\ref{ColourClump}, it can be seen that there is an inverse 
correlation between colour and clumpiness for both the observed galaxy 
sample and the MUGS simulated galaxy sample, although the anti-correlation of the MUGS galaxy sample is much weaker (and shifted in colour). The UW sample shows a positive correlation (though over a limited colour range) and the Dwarf sample shows no correlation. As 
the earlier-type disc galaxies are generally devoid of star forming 
regions and internal structure, and also tend to be red in colour, it 
follows that they would generally be located at the top left of 
Figure~\ref{ColourClump}, where the smoother, redder galaxies reside. 
Conversely, later-type galaxies tend to be bluer and contain high 
frequency residuals. For this reason, the anti-correlation is the expected correlation for Figure~\ref{ColourClump}.
 
The simulated and observed galaxies demonstrate a strong correlation between 
clumpiness and Hubble type. Due to the lack of redder galaxies, as with 
asymmetry, the simulated galaxies do not reach the top left corner of 
Figure~\ref{ColourClump}. However, unlike asymmetry, there are several 
galaxies from the MUGS sample that possess approximately zero 
clumpiness, consistent with their observed counterparts.

In Figure~\ref{ColourClump}, it can be seen that there is a group of 
galaxies (mainly from the Dwarf sample, with the exception of one MUGS 
and one UW galaxy), that are excessively clumpy. All the simulated 
galaxies that are excessively clumpy also have inclination angles of 
90$\degrees$ (see Table~\ref{tab:CAS}). Thus a study into the effects of 
dust attenuation on the clumpiness parameter was carried out. It was 
found that the clumpiness of the simulated galaxies was not affected by 
the contribution of the dust from \textsc{SUNRISE} \citep{Jonsson2006}. 
It was therefore concluded that the Dwarf sample, and Gal1 from the UW 
sample, are simply intrinsically clumpy.  In the case of the Dwarf 
sample, this is consistent with an analysis of the power spectrum of the 
cold gas associated with their interstellar media \citep{Pilkington2011}.
This may be attributed to the relatively low surface brightnesses 
associated with the dwarfs, as this causes the high frequency residuals 
to dominate in the calculation of the clumpiness.

\subsection{CAS comparison}

The CAS parameters are based solely on the morphology of galaxies, but 
are directly related to many intrinsic qualities. By comparing CAS 
parameters amongst themselves, as opposed to CAS versus colour, sampling 
biases, such as the colours of stellar populations, may be avoided. This 
allows for the structures of the galaxies to be compared directly, which 
may further emphasise the differences between observed and simulated 
galaxies.
 
\begin{figure}
\includegraphics[width=\hsize]{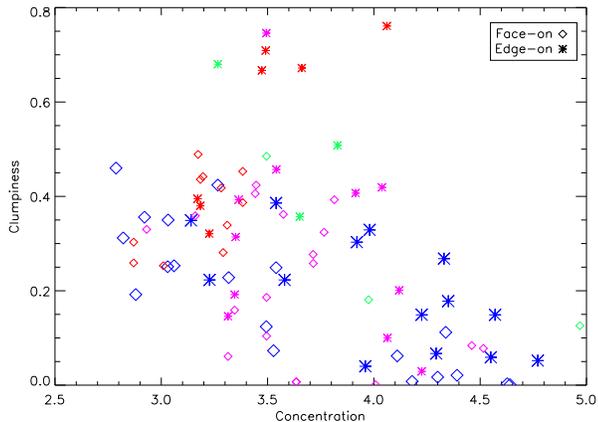}  
\small\caption{Relationship between the concentration (C) and clumpiness 
(S) parameters for both the observed and simulated galaxies. The points are 
coloured to represent the different galaxy samples. The \citet{Frei1996} 
sample of observed galaxies are coloured blue with larger symbols relative 
to the simulated samples. The MUGS sample are shown in purple, the UW 
sample in green, and the Dwarfs represented with red symbols. The 
symbols denote the inclination angle, as in Figure~\ref{AsymN}.}
\label{ConcClump}
\end{figure}

In Figure~\ref{ConcClump}, the observed galaxies from the \citet{Frei1996} 
sample demonstrate a general negative correlation between concentration 
and clumpiness. Galaxies that are smooth tend to be concentrated and 
those that are clumpy tend to be more diffuse. This correlation is 
generated because younger, disc galaxies tend to be both clumpy due to 
active star formation and diffuse, while older elliptical galaxies tend 
to be smoother and more dense.

The majority of simulated galaxies lie within the normal range of observed 
galaxies, in Figure~\ref{ConcClump}. It is apparent though that some of 
the simulations are excessively clumpy in comparison to their observed 
counterparts. As can be garnered from Table 1, all these simulated 
galaxies which are clumpy outliers only occur when viewed at an 
inclination of 90$\degrees$. This group of galaxies display 
approximately average concentration values, whilst being excessively 
clumpy, in comparison to both their observed and simulated galaxy 
counterparts. In regards to the excessively clumpy dwarfs, we note that 
the high frequency structure present in these simulated galaxies is 
consistent with the power spectrum analysis of the associated ISM by 
Pilkington et~al. (2011).

In Figure~\ref{ConcClump}, the MUGS sample demonstrates an approximately 
equivalent trend to the observed galaxy sample, despite their excessive 
bulge sizes. This is unexpected, as the central bulge component is 
disregarded in the computation of the clumpiness, but is included for 
the concentration parameter. However, with respect to the Dwarf and 
UW samples, the MUGS sample contains galaxies that are generally more 
concentrated. It is likely that the greater mass, velocity dispersion, 
and higher star formation threshold of the MUGS sample all contribute to 
both the larger bulge sizes and concentration of these galaxies, with 
respect to the other simulated galaxy samples. Furthermore, it is likely 
that these characteristics produced the earlier-type discs in the MUGS 
sample, which are expected to be less clumpy in nature.

\begin{figure}
\includegraphics[width=\hsize]{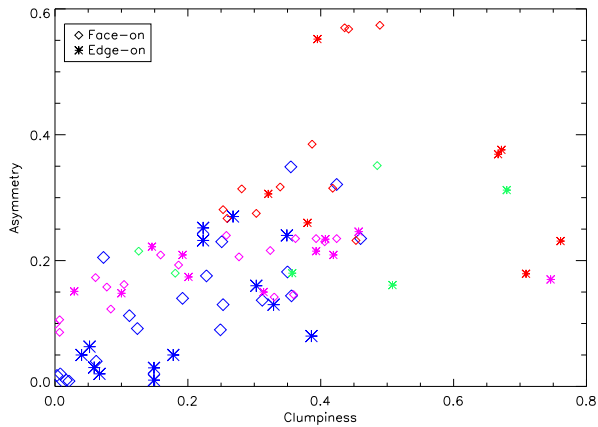} 
\small\caption{Relationship between clumpiness (S) and asymmetry (A) 
parameters for both observed and simulated galaxies. Symbols and colours 
as in Figure~\ref{ConcClump}.}
\label{ClumpAsym}
\end{figure}

In Figure~\ref{ClumpAsym}, it can be seen that there is a strong 
positive correlation between the clumpiness and asymmetry of observed 
galaxies. This correlation arises from the later-type galaxies tending 
to be clumpy and asymmetric, due to their star forming regions and 
intrinsic structure, and earlier-type galaxies tending to be devoid of 
structure and star forming regions, which pertains to them being more 
smooth and symmetric. The simulated galaxies appear to follow the same 
correlation. However, the sample of highly symmetric observed galaxies 
are not reproduced within the simulations.

In Figure~\ref{ClumpAsym}, the Dwarf sample of galaxies form three 
groups; those which lie within the normal range, those which are 
excessively clumpy, and those that are excessively asymmetric. As 
previously discussed, those galaxies that are excessively clumpy all 
have 90$\degrees$ inclination angles. Figure~\ref{ClumpAsym} shows that 
the high frequency spatial clumpiness of these galaxies is not directly 
related to asymmetry, as those galaxies that demonstrate excessive 
clumpiness possess a wide range of asymmetries with respect to other 
galaxies from the same sample.

The group of simulations which are seemingly highly asymmetric outliers 
at the top of Figure~\ref{ClumpAsym} is in fact just one simulated 
galaxy (DG2) viewed at various inclinations. This particular galaxy has 
a large star forming complex at $z=0$ which is offset from the centre of 
the galaxy (see Governato et al. 2010), explaining both the clumpiness 
and asymmetry of this galaxy.

\begin{figure}
\includegraphics[width=\hsize]{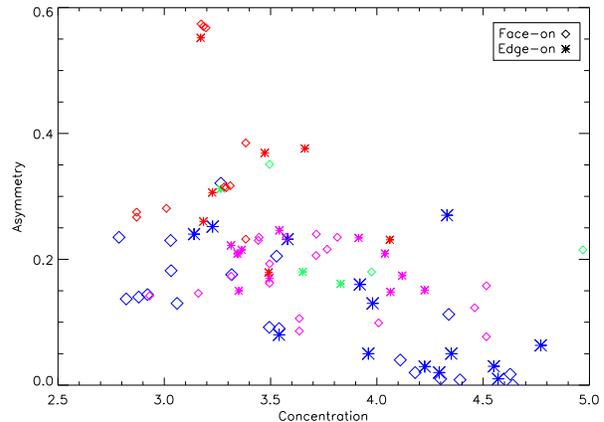}  
\small\caption{Relationship between concentration (C) and asymmetry (A) 
parameters for both observed and simulated galaxies. The points are coloured 
to represent the different galaxy samples. Symbols and colours as in 
Figure~\ref{ConcClump}.}
\label{ConcAsym}
\end{figure}

In Figure~\ref{ConcAsym}, there is an apparent negative correlation 
between asymmetry and concentration for both observed and simulated 
galaxies. This can be attributed to the fact that later-type galaxies 
are both diffuse and asymmetric, and earlier-type galaxies are generally 
dense with less internal structure. Within the Dwarf galaxy sample there 
is again a single galaxy viewed at various inclinations that 
demonstrates low concentration values coupled with extremely high 
asymmetry values. The large star forming complex in this simulation has 
not resulted in a concentration being outside the region defined by 
observed galaxies.

\section{Conclusion}
\label{Concl}

We have applied measures of concentration (C), asymmetry (A) and 
clumpiness (S) to a sample of simulated galaxies which have a range of 
masses and morphologies, as well as to a sample of observed galaxies. We 
have explored the correlations between these parameters as well as 
between each of them with B-V colour.  In general, reasonable agreement 
was found between the simulated and observed populations in these 
relations, although some differences become apparent, as summarised 
below.

Overall, the trend generated by the observed galaxy sample, with respect to 
colour (B-V), was approximately replicated by the simulated galaxies, 
although there is a distinct lack of redder simulated galaxies, which is 
related to a well-established inability of these simulations to halt 
late time star formation in early type galaxies \citep{Kawata2005}.

It was established that Hubble type is strongly linked with the three 
CAS parameters for the both the observed and simulated galaxy samples. This 
is expected as later-type galaxies, such as Hubble type Sc-Sd, have star 
forming regions, resulting in significant internal spatial structure, 
which manifests itself in significant asymmetries and clumpiness, and 
possess low concentration values. Conversely, earlier-type galaxies are 
smoother, more symmetric, and more concentrated, due to their relative 
lack of star forming regions, structure, and their evolutionary 
histories.

The concentration parameter was found to be robust and easily 
replicated. The values obtained for the simulated galaxies demonstrated 
an approximately equal range of values compared with their observed 
counterparts. The fewer number of simulated galaxies with high 
concentration merely reflected the relative paucity of early-type 
galaxies compared with the observed sample.

It was found that the range of asymmetries of the simulated galaxies did not extend to as low values as those of the most symmetric observed galaxies. This may be related to the ongoing star formation in the 
simulated galaxies, the same process which results in a dearth of red 
galaxies in the simulations. It is generally the early-type observed 
galaxies which are highly symmetric. Nevertheless, it remains a 
challenge for simulators to create highly-symmetric galaxies within a 
hierarchical structure formation paradigm.

A single dwarf galaxy simulation also shows excessive asymmetry compared 
to the observed sample. This is due to the existence of a large star 
forming complex offset from the centre of that simulation. Detailed 
exploration of the resolution of star forming complexes and star 
clusters within simulations is beyond the scope of this study, but it 
appears that the physical size of the star forming complex in at least 
one of the simulations is larger than observed, at least within the Frei 
a et al. (1996) sample.

As it is a well known issue that simulated galaxies have 
disproportionately large bulge components \citep{Stinson2010}, it was 
decided that a central region, pertaining to the size of the bulge 
component, would be removed for all galaxies in the calculation of the 
clumpiness. Following this, we found that the simulated and observed 
galaxies showed similar spreads in the values of clumpiness. The 
early-type disc galaxies from the MUGS sample, which have lower rates 
star formation occurring in the disk at $z=0$, have low S, while the 
later-type simulated galaxies continue to have high levels of star 
formation, and high values of S, consistent with their observed 
counterparts. Analysis of the clumpiness values generated by the 
simulated galaxies also demonstrated that several simulated galaxies 
were excessively clumpy when viewed at an inclination angle of 
90$\degrees$. At least in relation to the dwarfs, this is likely related 
to high frequency power of these galaxies, when analysed at an edge-on 
inclination, a characteristic that is not consistent with observations 
(cf., Pilkington et~al. 2011, for the high-frequency analysis of the ISM 
of these dwarfs).

\section*{Acknowledgments}

BKG, CBB, and CJC acknowledge the support of the UK's Science \& 
Technology Facilities Council (ST/F002432/1, ST/G003025/1). This work 
was made possible by the University of Central Lancashire's High 
Performance Computing Facility, the UK's National Cosmology 
Supercomputer (COSMOS), NASA's Advanced Supercomputing Division, 
TeraGrid, the Arctic Region Supercomputing Center, the Shared Hierarchical Academic Research Computing Network (SHARCNET) and the University of 
Washington. We thank the DEISA consortium, co-funded through EU FP6 
project RI-031513 and the FP7 project RI-222919, for support within the 
DEISA Extreme Computing Imitative. We would also like to thank Elisa House for the generation of the observed and simulated galaxy images in Figures~1-4.


\bibliographystyle{mn2e} 
\bibliography{CasRef}

\appendix
\section{Computation of the High Spatial Frequency Clumpiness Parameter (S)}

For the purpose of comparing simulated galaxies with those observed, the 
central bulge component of each galaxy was removed in the computation of 
the high frequency spatial clumpiness. However, the method proposed by 
\citet{Conselice2003} favoured the removal of a central region, 
amounting to 0.05 $\times$ the Petrosian radius ($r_{\rm Petrosian}$), 
for the purpose of excluding any unresolved regions of the image only. 
As the measure of high spatial frequency clumpiness specified by 
\citet{Conselice2003} forms an integral part of the CAS system, the 
method of computation is presented here.

Initially the image is sky-subtracted, flat-fielded, and the 
characteristic radius determined. The radius is defined as 1.5 $\times$ 
$r_{\rm Petrosian}$ (further details can be found in \citet{Bershady2000}). If the data is undersampled, the original image undergoes a 
boxcar smooth of 5x5 pixels prior to any further computation; this image 
is then re-defined as the original image. The next step involves 
smoothing the image by a factor $r_{\rm Petrosian}$/6, again using a 
boxcar smooth. Following this, a map of the residuals is created by 
subtracting the smoothed image from the original image. Within this 
residual image, all the negative pixels, and the pixels within the 
central region of 0.05 $\times$ $r_{\rm Petrosian}$, are then forced to 
zero, leaving behind the high frequency residuals only. The total flux 
from the residual map, within the specified radius, is then summed and 
divided by the total flux of the initial galaxy, within the specified 
radius. The value generated equates to the clumpiness of the galaxy 
including the clumpiness in the background.

To ensure that the results of the clumpiness parameter determination are 
unaffected by noise, further background correction is then required. 
This involves selecting an area from the image that contains background 
only and repeating the aforementioned procedure. Note that the 
background should also be smoothed by a boxcar smooth of 5$\times$5, if 
the data is undersampled. Once selected, the area should be smoothed by 
a boxcar smooth of $r_{\rm Petrosian}$/6, using the value of the 
Petrosian radius obtained from the galaxy image. A background residual 
map is then created by subtracting the smoothed background image from 
the original background image and forcing the negative pixels to zero. 
The background flux is then be summed and normalised to the the number 
of pixels that constitute the total galaxy, within the specified radius. 
This value is then divided by the summation of the total flux of the 
galaxy thus producing the clumpiness in the background. The clumpiness 
in the background is then subtracted from the aforementioned clumpiness 
in the galaxy, which is computationally defined in \S\ref{Clump}.

\label{lastpage}
\end{document}